\documentstyle{article}

\begin{document}

\begin{center}
{\Large\bf
Scattering of scalar and Dirac particles \\ by a magnetic tube of finite radius}
\\[1cm]
{\bf  Vladimir D. Skarzhinsky$^{\dagger, \ddagger,}$\footnote{{\it e-mail:
vdskarzh@sgi.lpi.msk.su}} and
J\"urgen Audretsch$^{\dagger,}$\footnote{{\it e-mail:
Juergen.Audretsch@uni-konstanz.de}}}
\\[0.5cm]
$^\dagger$Fakult\"at f\"ur Physik der Universit\"at Konstanz,
Postfach 5560 M 673,\\ D-78434 Konstanz, Germany \\[0.5cm]
$^\ddagger$P.~N.~Lebedev Physical Institute, Leninsky prospect 53,
Moscow 117924, Russia
\\[1cm]

\begin{abstract}
\noindent
We consider the Dirac equation in cylindrically symmetric magnetic fields and find its normal modes as eigenfunctions of a complete set of commuting operators. This set consists of the Dirac operator itself, the $z$-components of the linear and the total angular momenta, and of one of the possible spin polarization operators. The spin structure of the solution is completely fixed independently of the radial distribution of the magnetic field which influences only the radial modes.

We solve explicitly the radial equations for the uniform magnetic field inside a solenoid of a finite radius and consider in detail the scattering of scalar and Dirac particles in this field. For particles with low energy the scattering cross section coincides with the Aharonov--Bohm scattering cross section. We work out the first order corrections to this result caused by the fact that the solenoid radius is finite. At high energies we obtain the classical result for the scattering cross section.


\end{abstract}
\end{center}



\section{Introduction}\label{intro}

The behavior of relativistic charged particles in external magnetic fields has been the subject of many investigations in QED (see, for example, \cite{Akhiezer65,Landau79}). Most of them has been concerned with the  synchrotron radiation emitted by charged particles moving in cyclic accelerators and storage rings \cite{Sokolov68}. This effect has found wide applications in different fields of physics, biology and technology.

By its origin the synchrotron radiation is the classical effect. It has been shown that quantum corrections  became important only for relativistic particles or intense magnetic fields \cite{Sokolov68,Klepikov54}. This happens at high energies $E$ and high field strengths $B$ leading to a characteristic product \cite{Schwinger51}
\begin{equation}\label{Schwinger}
{B\over B_0}{E\over Mc^2} \approx 1,\quad {\rm with} \quad B_0 =
{M^2c^3\over e\hbar}
\end{equation}
of the order 1. Other QED processes, such as pair production and pair annihilation, have no classical counterparts. In these cases too, intensities of the processes become relevant if particle energies and magnetic fields fulfil Eq.(\ref{Schwinger}). Therefore, in QED processes, the external magnetic field can not be treated as a perturbation. The relativistic particles must instead be described by exact solutions of the Dirac equation in an external magnetic field. This will also be done in this paper. For the comprehensive review of exact solutions of relativistic wave equations in external fields see \cite{Bagrov90}.

Much work has been devoted to solutions of the Dirac equation in an uniform electric and magnetic fields \cite{Erber66,Ritus79} and in the electromagnetic plane wave field \cite{Volkov35}. Different QED processes have been studied for the corresponding physical situations \cite{Erber66,Ritus79}.

A new field of research was initiated with the study of the Aharonov--Bohm (AB) effect (for the latter see \cite{Aharonov59,Peshkin89}). QED processes in the presence of a magnetic string were elaborated in detail; the differential cross sections of the bremsstrahlung of an electron passing by the magnetic string \cite{Audretsch96} and the pair production by a single photon in this potential \cite{Skarzhinsky96} have been worked out. In these cases it surprisingly turned out that the cross sections do not become small even for low particle energies. This fact can be interpreted as follows: for a finite flux the string magnetic field becomes infinitely strong in the limit of a string, so that the criterion of Eq.(\ref{Schwinger}) is formally fulfilled for all energies.

To describe realistic situations the magnetic string has to be  replaced by a thin solenoid which contains an intense but finite magnetic field. It is now very important to know how the results obtained in the string limiting case are modified for a solenoid of finite radius. We will attack this problem in this and following papers where we are going to investigate in detail different QED processes in cylindrical magnetic fields. Below we concentrate on the scattering of scalar and Dirac particles.

In the Sections 2 and 3 we define the normal modes of the Dirac equation in cylindrical magnetic field of arbitrary radial dependence as eigenfunctions of a complete set of commuting operators. The set consists of the Dirac operator itself, the $z$-components of linear and total angular momenta, and also, of one of the possible spin polarization operators. We fix completely the spin structure of the solution independently of the radial distribution of the magnetic field which influences only the radial modes. In Sec.4 we solve explicitly the radial equations for the model of a solenoid of finite radius with an uniform magnetic field. With the tube radius going to zero this model can be considered as a realistic model replacing the AB magnetic string. This approach allows to solve correctly the Dirac equation for the AB potential. Then we consider in detail (Sec.5) the scattering of scalar and Dirac particles by the magnetic field of the solenoid of the finite radius. At low energies of incident particles the scattering cross section coincides with the AB scattering cross section; we derive the first order corrections caused by the finite tube radius. At high energies we obtain the classical result for the scattering cross section.

Throughout we use units such that $\hbar=c=1.$

\section{The Dirac equation in a cylindrically symmetric magnetic field}\label{dirac}

The Dirac equation in an external magnetic field reads in cylindrical coordinates $(\rho,\;\varphi,\;z)$
\begin{equation}\label{de}
i \partial_{t}\Psi(x) = H\Psi(x),\quad H = \alpha_{i}(\hat p_{i} -
eA_{i}) + \beta M
\end{equation}
where $x = (t, \rho, \varphi, z)$, $e$ is the charge of the Dirac particle ($e>0$ for the positron and $e<0$ for the electron). $\alpha_i$ and $\beta$ are the known matrices written in the cylindrical coordonates, and the kinetic momenta are given by
\begin{eqnarray}
\hat\pi_{\rho} : &=& \hat p_{\rho} = - i\partial_{\rho},\quad \hat p_{3} : = -i \partial_{z}\,, \nonumber \\
\quad \hat\pi_{\varphi}: &=& \hat p_{\varphi} - eA_{\varphi}=-{i \over
\rho}\partial_{\varphi}-eA_{\varphi}\,.
\end{eqnarray}

The vector potential for an {\sl arbitrary} cylindrically symmetric magnetic field of a fixed direction (along $z$-axis) has a nonzero angular component $A_{\varphi}(\rho),$ and the magnetic field reads in terms of the potential
\begin{equation}\label{B}
B_z(\rho)=A'_{\varphi}(\rho)+{1\over\rho}\;A_{\varphi}(\rho)\,.
\end{equation}

The dependence of the solutions on $z$ and the azimuthal angle $\varphi$ can be fixed in demanding that $\Psi(x)$ is an eigenfunction of the operators of linear momentum projection $\hat{p_3}$ and total angular momentum projection $\hat{J_3},$
\begin{eqnarray}\label{co}
\hat{p_3}\Psi(x) &=& p_3\Psi(x) \,,\nonumber \\
\hat{J_3}\Psi(x) &=& (\hat L_3+{1 \over 2} \Sigma_3)\Psi(x)
= (-i\partial_{\varphi}+{1 \over 2} \Sigma_3)\Psi(x)
= j_3 \Psi(x) \,,\quad j_3 : = l+{1\over2}.
\end{eqnarray}
Here $l$ is the integral part of the half-integral eigenvalue $j_3.$

We find
\begin{equation}\label{ds}
\Psi(x) = {1\over 2\pi} e^{-iE_p t + ip_3z}\;\psi(\rho, \varphi),
\quad \psi(\rho, \varphi) =  \pmatrix{ u \cr v\cr},
\end{equation}
where
\begin{equation}\label{uv}
u =\left(\begin{array}{c}
\displaystyle
C_1\;R_1(\rho)\;e^{ il\varphi}\\
\displaystyle
C_2\;R_2(\rho)\;e^{i(l+1)\varphi}
\end{array}\right), \;
v =\left(\begin{array}{c}
\displaystyle
C_3\;R_3(\rho)\;e^{il\varphi}\\
\displaystyle
C_4\;R_4(\rho)\;e^{i(l+1)\varphi}
\end{array}\right),
\end{equation}
$E_p = \sqrt{p^2+M^2} = \sqrt{p_\perp^2 + p_3^2 +M^2}$ is the energy, $p_\perp$ denotes the radial momentum and $C_i$ are the spin polarization coefficients.

In terms of $u-$ and $v-$spinors the Dirac equation reads
\begin{eqnarray}\label{uve}
\sigma_i(p_i-eA_i)\;u = (E_p+M)v\,,\nonumber \\
\sigma_i(p_i-eA_i)\;v = (E_p-M)u\,.
\end{eqnarray}
From Eqs.(\ref{uve}) and (\ref{uv}) we find equations for the radial functions $R_i,$
\begin{eqnarray}\label{re}
-iC_2 R_2'-i\left({l+1\over\rho}-eA_{\varphi}\right) C_2 R_2+p_3 C_1 R_1 &=& (E_p+M)C_3 R_3  \,,\nonumber \\
-iC_1 R_1'+i\left({l\over\rho}-eA_{\varphi}\right) C_1 R_1-p_3 C_2 R_2  &=& (E_p+M)C_4 R_4 \,,\nonumber \\
-iC_4 R_4'-i\left({l+1\over\rho}-eA_{\varphi}\right) C_4R_4+p_3 C_3R_3  &=& (E_p-M)C_1 R_1 \,,\nonumber \\
-iC_3 R_3'+i\left({l\over\rho}-eA_{\varphi}\right)C_3 R_3-p_3 C_4 R_4  &=& (E_p-M)C_2 R_2 \,.
\end{eqnarray}
The spin coefficients can be constrained by the normalization condition
\begin{equation}\label{|C|}
\sum_{i=1}^{i=4}|C_i|^2 = 1.
\end{equation}

\section{Spin polarization states}\label{spin}

In the case of cylindrical magnetic fields the spin polarization coefficients $C_i$ can be fixed by one of two spin operators $\hat S_t$ or $\hat S_3$ \cite{Sokolov68} which commutes with operators $\hat{H},\hat{p_3}$ and $\hat{J_3}$ but do not commute with each other. The helicity operator,
\begin{equation}\label{st}
\hat S_t : = {\vec{\Sigma}\cdot(\hat{\vec{p}}-e\vec{A})\over p},
\end{equation}
describes longitudinal polarization (the projection of the spin onto the velocity direction of the Dirac particle). The $z$-component $S_3$ of the operator $\hat{\vec{S}}: =\beta\;\vec{\Sigma} +\gamma (\hat{\vec{p}}-e\vec{A})/M\,,$
\begin{equation}\label{s3}
\hat S_3 : =\beta\;\Sigma_3 + \gamma \;{\hat{p}_3\over M}, \quad
\gamma = \pmatrix{      0   &  1\cr
                        1   &  0 \cr}
\end{equation}
defines a transverse polarization (along the direction of the magnetic field) for nonrelativistic motion, or for the motion in the plane perpendicular to the magnetic field. The operators $\hat S_t$ and $\hat S_3$ are not independent. However, to make things easy for the reader we will discuss them separately.

\subsection{The helicity states}\label{hel}

For the solution of the Dirac equation (\ref{de}) which is the eigenstate of the helicity operator (\ref{st}),
\begin{equation}\label{st1}
\hat S_t \Psi(x) = s \Psi(x), \quad s=\pm 1,
\end{equation}
the Eqs.(\ref{uve}) and (\ref{st}) connect the $u-$ and $v-$spinors according to
\begin{equation}\label{vu'}
v = s\;{\sqrt{E_p-M}\over\sqrt{E_p+M}}\;u\,.
\end{equation}
It means that $R_3=R_1, R_4=R_2$ and that the coefficients $C_3, C_4$ are coupled by Eq. (\ref{vu'}) to $C_1, C_2$ correspondingly. Choosing $C_1$ and $C_2$ to be
\begin{equation}\label{c1c2'}
C_1 = {\sqrt{E_p+M}\sqrt{p+sp_3}\over \sqrt{2E_p}\sqrt{2p}}, \; C_2
= {is\sqrt{E_p+M}\sqrt{p-sp_3}\over \sqrt{2E_p}\sqrt{2p}}
\end{equation}
and determining coefficients $C_3$ and $C_4$ from Eq.(\ref{vu'}),
\begin{equation}\label{c3c4'}
C_3 = {s\sqrt{E_p-M}\sqrt{p+sp_3}\over \sqrt{2E_p}\sqrt{2p}}, \;
C_4 = {i\sqrt{E_p-M}\sqrt{p-sp_3}\over \sqrt{2E_p}\sqrt{2p}}\,,
\end{equation}
we obtain from Eq.(\ref{re}) the following set of the equations for the independent radial components $R_{1,2},$
\begin{eqnarray}\label{rest}
R_2'+\left({l+1\over\rho}-eA_{\varphi}\right) R_2 &=& p_\perp R_1
\,,\nonumber \\
-R_1'+\left({l\over\rho}-eA_{\varphi}\right) R_1 &=& p_\perp R_2  \,.
\end{eqnarray}
The solution of the Dirac equation with the quantum numbers $E, p_3, j_3, s$ has then the form of Eqs.(\ref{ds}), (\ref{uv}) with $R_3=R_1, R_4=R_2$ and the coefficients (\ref{c1c2'}), (\ref{c3c4'}).

\subsection{The transverse polarization states}\label{tran}

On the other hand, for the eigenstate of the operator $\hat S_3$ (\ref{s3}),
\begin{equation}\label{s31}
\hat S_3 \Psi(x) =s_3 \Psi(x), \quad s_3=s\,\sqrt{1+{p_3^2\over M^2}}, \quad
s=\pm 1,
\end{equation}
we find from Eqs.(\ref{uve}) and (\ref{s3}) that
\begin{equation}\label{vu''}
v_1 = {p_3\over M(s_3+1)}\;u_1, \quad v_2 = {p_3\over M(s_3-1)}\;u_2\,.
\end{equation}
This means that $R_3=R_1, R_4=R_2$ and that the coefficients $C_3, C_4$ are defined by (\ref{vu''}) from $C_1, C_2,$ correspondingly. Fixing the coefficients $C_1$ and $C_2$ by the expressions
\begin{equation}\label{c1c2''}
C_1 = \sqrt{s_3+1\over 2s_3}\;{\sqrt{E_p+s_3M}\over \sqrt{2E_p}}, \quad
C_2 = i\epsilon (s_3p_3)\sqrt{s_3-1\over 2s_3}\;{\sqrt{E_p-s_3M}\over \sqrt{2E_p}}
\end{equation}
where $\epsilon (x)$ is the sign of $x$ and determining the coefficients the $C_3$ and $C_4$ from Eq.(\ref{vu''}),
\begin{equation}\label{c3c4''}
C_3 = \epsilon (s_3p_3)\sqrt{s_3-1\over 2s_3}\;{\sqrt{E_p+s_3M}\over
\sqrt{2E_p}}, \quad
C_4 = i\sqrt{s_3+1\over 2s_3}\;{\sqrt{E_p-s_3M}\over\sqrt{2E_p}}\,,
\end{equation}
we obtain from Eq.(\ref{re}) again the radial equations (\ref{rest}).
The solution of the Dirac equation with the quantum numbers $E, p_3, j_3, s_3$ has therefore the form Eqs.(\ref{ds}), (\ref{uv}) with $R_3=R_1, R_4=R_2$ and the coefficients (\ref{c1c2''}), (\ref{c3c4''}).

We have shown how the spin polarization operators (\ref{st}) or (\ref{s3}) fix  the spin structure of the solutions of the Dirac equation in arbitrary cylindrical magnetic field. Note that solutions for different configurations of the magnetic field differ from each other only by their radial functions.

\subsection{The charge conjugate states}\label{charge}

The solutions (\ref{ds}) of the Dirac equation (\ref{de}) represent electron ($e<0$) and positron ($e>0$) wave functions of positive energy, $\Psi_{e,p}(x)$. The complete set of solutions of the Dirac equation (\ref{de}) includes the positive and negative energy electron (or positron) states. Instead of negative energy electron (or positron) states one can use as well the charge conjugated positron (electron) states which can be obtained by the charge conjugation operation,
\begin{equation}\label{conj}
\Psi(x) \rightarrow \Psi_c(x) : = C\overline{\Psi}_{\rm transp}(x), \quad
C=\alpha_2.
\end{equation}
For the helicity and transverse states we obtain
\begin{equation} \label{conjds}
\Psi_c(x) = {1\over 2\pi}\;e^{iE_pt-ip_3z}\;\psi_c(\rho,\varphi),
\; \psi_c(\rho,\varphi) = \pmatrix{ u_c \cr v_c\cr}
\end{equation}
where
\begin{eqnarray}\label{conjuv}
u_c = \left(\begin{array}{c}\displaystyle
iC_4^*\;R^*_2(\rho)\;e^{-i(l+1)\varphi}\\
\displaystyle -iC_3^*\;R^*_1(\rho)
\;e^{-il\varphi}\end{array}\right) \,, \nonumber \\
v_c = \left(\begin{array}{c}\displaystyle -iC_2^*\;R^*_2(\rho)
\;e^{-i(l+1)\varphi}\\
\displaystyle iC_1^*\;R^*_1(\rho)\;e^{-il\varphi}
\end{array}\right)
\end{eqnarray}
with the coefficients (\ref{c1c2'}), (\ref{c3c4'}) and (\ref{c1c2''}), (\ref{c3c4''}), correspondingly.

With reference to these states, the electron-positron field operator reads
\begin{equation}\label{epfo}
\hat{\psi}(x) = \sum_j [\Psi_e(x)\;a_j + \Psi_p^c(x)\;b_j^{\dagger}]
\end{equation}
where $a_j$ and $b_j$ are the annihilation operators for electrons and positrons with quantum numbers $j := (p_\perp, p_3, l, s).$ This field operator obeys Eq.(\ref{de}) with $e<0.$ Of course, we could use the charged conjugated field operator which obeys this equation with $e>0.$

\subsection{The radial equations}\label{rad}

The set of the radial equations (\ref{rest}) is equivalent to the following second order equation for the first component
\begin{equation}\label{reR1}
R_1''+{1\over\rho} R_1' - \left[\left({l\over\rho} - eA_{\varphi}\right)^2  -  e\left(A_{\varphi}'+{1\over\rho}A_{\varphi}\right) - p_\perp^2\right]R_1 = 0;
\end{equation}
the second component is then defined by the equation
\begin{equation}\label{R2}
R_2 = {1\over p_\perp }\left[-R'_1 +
\left({l\over\rho}-eA_{\varphi}\right)R_1 \right].
\end{equation}
Alternatively one could start with the second order equation for $R_2$ and define the first component $R_1$ by the equation (\ref{rest}). Both ways are completely equivalent. We would like to stress that because of the first order constraints (\ref{rest}) between the components $R_1$ and $R_2,$ the Dirac equation differs from the Klein--Gordon equation for relativistic scalar particles.

The radial functions can be normalized by the condition
\begin{equation}\label{Rnc}
\int \rho d\rho\;R^*_{1,2}(p'_\perp\rho)\;R_{1,2}(p_\perp \rho) =
{\delta(p_\perp-p'_\perp)\over\sqrt{p_\perp p'_\perp}}.
\end{equation}
The Eqs. (\ref{reR1}), (\ref{R2}) together with the normalization condition (\ref{Rnc}) define the radial functions $R_1$ and $R_2$ up to a common phase factor.

\bigskip

{\bf The nonrelativistic limit:}
The positron and electron radial and angular functions conserve their forms in the nonrelativistic approximation ($E_p=M+{\cal E}_p,\;{\cal E}_p\ll M$) with nonzero spin coefficients
\begin{eqnarray} \label{nps}
C_1 &=& {\sqrt{p+sp_3}\over \sqrt{2p}}, \quad C_2 = {\sqrt{p-sp_3}\over
\sqrt{2p}},\quad \mbox {for helicity states} \nonumber  \\
C_1 &=& {\sqrt{s+1}\over \sqrt{2s}}, \quad C_2 = {\sqrt{s-1}\over
\sqrt{2s}}, \quad \mbox {for transverse states}
\end{eqnarray}

\section{Solenoid of a finite radius and the magnetic string} \label{models}

Our basic aim is to study the AB scattering of scalar and Dirac particles for experimentally realistic situations. We consider therefore a solenoid of finite radius $\rho_0$ with the uniform magnetic field inside. This model contains the AB magnetic string in the limit of zero radius and constant magnetic flux. This limit can alternativly be based on a different model \cite{Hagen} for which the magnetic field is concentrated on the surface of a cylinder of radius $\rho_0.$ It is obvious that this second model being more simple for the investigation of zero radius limit does not describe a realistic experimental set up.

\subsection{The uniform magnetic field inside a solenoid of finite radius}\label{finrad}

The uniform magnetic field $(0, 0, B)$ restricted to the interior of a solenoid of finite radius $\rho_0$ can be described by the vector potential
\begin{equation} \label{solvp}
eA_{\varphi} = \cases{\;\; {eB\rho/2} & \cr {e\Phi/2\pi\rho} & \cr} =
\cases{\;{\phi \rho /\rho_0^2} &\quad if $\;\rho\leq \rho_0$ \cr
\;{\phi/\rho} &\quad if $\;\rho\geq \rho_0$ \cr}
\end{equation}
where $\Phi=\pi\rho_0^2 B := \beta\Phi_{0}=\varepsilon\phi\Phi_{0}$ is the magnetic flux in the solenoid, $\Phi_{0} := 2\pi /|e|$ is the magnetic flux quantum and $\varepsilon$ is the sign of the  charge ($\varepsilon>0$ for positrons and $\varepsilon<0$ for electrons). We take $\beta>0$ so that the constant magnetic field inside the solenoid
\begin{equation}\label{solB}
B={2\beta\over |e|\rho_0^2}, \quad \rho\leq\rho_0
\end{equation}
points in the positive $z$ direction. To obtain the uniform magnetic field in the whole space one may perform the limiting procedure $\beta\rightarrow |e|B\rho_0^2/2$ together with $\rho_0\rightarrow\infty.$

Let us consider first the internal solution. The general radial equations (\ref{reR1}) and (\ref{R2}) read in this case
\begin{eqnarray} \label{reR1'}
R_1''+{1\over\rho} R_1'&-&\left[\left({l\over\rho}-
{\phi\over\rho_0^2}\;\rho\right)^2-2{\phi\over\rho_0^2}-
p_\perp^2\right]R_1 = 0\,, \\
\label{R2'}
R_2 &=& {1\over p_\perp }\left[-R'_1 +
\left({l\over\rho}-{\phi\over\rho_0^2}\;\rho\right)R_1 \right].
\end{eqnarray}
Introducing the new variable $x=\beta (\rho^2/\rho_0^2)$ and the new functions $Y_{1,2}(x)$ according to
\begin{equation}\label{Ry}
R_1(x) = x^{|l|\over 2}\;e^{-{x\over 2}}\;Y_1(x)\,, \quad
R_2(x) = x^{|l+1|\over 2}\;e^{-{x\over 2}}\;Y_2(x)
\end{equation}
we obtain from Eq.(\ref{reR1'}) the confluent hypergeometric equation \cite{Gradshteyn80}
\begin{eqnarray}\label{y1}
&&xY_1''+(B_1-x)Y_1'-A_1Y_1=0\,, \\ \nonumber
&&A_1={|l|-\varepsilon l+1-\varepsilon\over 2} -{p_\perp^2 \rho_0^2\over 4\beta}\,, \quad
B_1=|l|+1\,.
\end{eqnarray}
Then from Eqs.(\ref{R2'}) - (\ref{y1}) we find the following confluent hypergeometric equation for the function $Y_2(x)$
\begin{eqnarray}\label{y2}
&&xY_2''+(B_2-x)Y_2'-A_2Y_2=0\,, \\ \nonumber
&&A_2={|l+1|-\varepsilon(l+1)+1+\varepsilon\over 2}-{p_\perp^2 \rho_0^2\over 4\beta}\,, \quad  B_2=|l+1|+1\,.
\end{eqnarray}
Solutions of Eqs.(\ref{y1}) and (\ref{y2}) which are regular at $\rho=0$ are the confluent hypergeometric functions $\Phi(A_1,B_1;x)$ and $\Phi(A_2,B_2;x),$ correspondingly, so that we obtain
\begin{eqnarray}\label{RF}
R_1^{\rm int}(x)&=&c_l\;x^{|l|\over 2}\;e^{-{x\over 2}}\;\Phi(A_1,B_1\,;x)\,,
\nonumber \\
R_2^{\rm int}(x)&=&\tilde{c}_l\;x^{|l+1|\over 2}\;e^{-{x\over 2}}\;\Phi(A_2,B_2\,;x)\,.
\end{eqnarray}
The coefficients $c_l$ and $\tilde{c}_l$ are connected with each other by the Eq.(\ref{R2'}).

The external solution obeys the radial equations (\ref{reR1}) and (\ref{R2'}) at $\rho>\rho_0.$
\begin{eqnarray} \label{Bess}
&&R_1''+{1\over\rho} R_1'-{(l-\phi)^2\over\rho^2}\;R_1 + p_\perp^2\;R_1 = 0\,, \nonumber \\
&&R_2 = {1\over p_\perp}\left(-R_1'+ {l-\phi \over\rho} R_1\right)\,.
\end{eqnarray}
Solving these equations by Bessel functions of the first kind with positive and negative order and using Eq.(\ref{R2}) we find the external radial components
\begin{eqnarray} \label{Rs}
R^{\rm ext}_1(\rho) &=& a_l\;J_{\nu}(p_\perp \rho) + b_l\;J_{-\nu}(p_\perp \rho)\,,
\nonumber \\
R^{\rm ext}_2(\rho) &=& \epsilon_{l-\phi}\left[a_l\;J_{\tilde\nu}(p_\perp \rho) - b_l\;J_{-\tilde\nu}(p_\perp \rho)\right]
\end{eqnarray}
where
\begin{equation} \label{order1}
\nu : = |l-\phi|\,, \quad \tilde\nu : =\nu+\epsilon_{l-\phi}\,, \;
\epsilon_{l-\phi} : = \left\{\begin{array}{rl}
1 &\; {\rm if}\ l >\phi\\
-1 &\; {\rm if}\ l<\phi
\end{array}\right..
\end{equation}

The coefficients $a_l$ and $b_l$ can be obtained from the matching conditions at the surface $\rho = \rho_0$ or $(x = \beta)$ of the solenoid for the internal and external solutions. Since radial components are connected with each other by Eq.(\ref{R2'}) one can use any couple of the matching conditions for components $R_1, R_2$ and their first derivatives $R'_1, R'_2$ . Choosing the conditions for $R_1$ and $R'_1$ we obtain
\begin{eqnarray}\label{mc1}
&&c_l\beta^{|l|\over 2}\;e^{-{\beta\over
2}}\;\Phi(A_1,B_1;\beta) = a_l\;J_{\nu}(p_\perp \rho_0) +
b_l\;J_{-\nu}(p_\perp \rho_0)\,,  \\
&&{2\beta\over p_\perp \rho_0}
\left[{|l|-\beta\over2\beta}+{\Phi'(A_1,B_1;\beta)
\over\Phi(A_1,B_1;\beta)}\right] c_l\beta^{|l|\over
2} e^{-{\beta\over 2}} \Phi(A_1,B_1;\beta) = a_l J'_{\nu}(p_\perp
\rho_0) + b_l J'_{-\nu}(p_\perp \rho_0). \nonumber
\end{eqnarray}
Eliminating the coefficient $c_l$ from these equations we find the matching conditions in the form
\begin{eqnarray}\label{mc2}
&&{2\beta\over p_\perp
\rho_0}\left[{|l|-\beta\over2\beta}+{\Phi'(A_1,B_1;\beta)
\over\Phi(A_1,B_1;\beta)}\right] =  \\
&&{2\beta\over p_\perp \rho_0}\left[{|l|-\beta\over2\beta}+{A_1\over
B_1}{\Phi(A_1+1,B_1+1;\beta) \over\Phi(A_1,B_1;\beta)}\right] =
{a_l\;J'_{\nu}(p_\perp \rho_0) + b_l\;J'_{-\nu}(p_\perp \rho_0)\over
a_l\;J_{\nu}(p_\perp \rho_0) + b_l\;J_{-\nu}(p_\perp \rho_0)} \nonumber
\end{eqnarray}
which fixes all coefficients up to a normalization constant. The same results can be obtained for instance from the continuity conditions for $R_1$ and $R_2$ at $\rho=\rho_0.$

\subsection{The zero radius limit (magnetic string)}\label{zerorad}

The vector potential for the infinitely thin, infinitely long straight magnetic string lying along $z$-axis
\begin{equation} \label{abp}
eA_{\varphi} = {e\Phi \over 2\pi\rho} = {\varepsilon\Phi \over
\Phi_{0}\rho} = {\varepsilon\beta\over\rho}= {\phi\over\rho}\,, \quad 0\leq\rho<\infty
\end{equation}
can be obtained from the vector potential of the solenoid (\ref{solvp}) for  $\rho_0\rightarrow 0$ keeping the magnetic flux constant. It is singular on $z$-axis and produces the singular magnetic field
\begin{equation} \label{abB}
B = {2\beta\over |e|\rho} \delta(\rho) \,,
\end{equation}
which is concentrated on the $z$-axis and points to the positive $z$ direction. We separate the integral number $N$ from the flux parameter $\beta$, $\beta=N+\delta,\;N\geq 0,\; 0\leq\delta<1$ since it is the fractional part $\delta$ of the dimensionless magnetic flux which produces all physical effects. Its integral part $N$ will appear as a phase factor $\exp(i\varepsilon N\varphi)$ in the solutions to the Dirac equation.

In the case of the AB potential the radial solutions are given by Eqs.(\ref{Rs}), but have another domain of definition. We note that the component $R_1$ contains modes with coefficients $b_l$ which are singular at $\rho = 0$ for all $l.$ The component $R_2$ contains singular modes with coefficients $a_N\;(\varepsilon>0),\;a_{-N-1}\;(\varepsilon<0)$ and $b_{l\neq N}\;(\varepsilon>0),\;b_{l\neq -N-1}\;(\varepsilon<0).$ The appearance of the singular modes is usually forbidden by the normalization condition (\ref{Rnc}) for radial functions. But for unbounded potentials, and for the AB potential (\ref{abp}) in particular, this condition is less restrictive. The normalization condition (\ref{Rnc}) requires that the integrals of the type
\begin{equation}\label{I}
I_{\mu} = \int_0^{\infty} J_{\mu}^2(p_\perp \rho) \rho d\rho
\end{equation}
were convergent at small $\rho$ what takes place at $\mu > - 1.$ For the positron solution $(\varepsilon>0)$ the normalization condition (\ref{I}) for $R_1$ allows nonzero coefficients $a_l,\;b_N$ and $b_{N+1}$ while this condition for $R_2$ allows nonzero coefficients $a_l,\;b_N$ and $b_{N-1}.$ For the Dirac equation both components $R_1$ and $R_2$ must satisfy the normalization condition (\ref{I}). It means that only the coefficient $b_N$ can be nonzero presenting one singular mode with $l=N$ in the positron solution of the Dirac equation (while two singular modes are allowed for scalar wave equations). For the electron solution $(\varepsilon<0)$ the normalization condition (\ref{I}) for $R_1$ allows nonzero coefficients $a_l,\;b_{-N}$ and $b_{-N-1}$ while the condition for $R_2$ allows nonzero coefficients $a_l,\;b_{-N-1}$ and $b_{-N-2}.$ Therefore only one singular mode with $l=-N-1$ can be present in the electron solution of the Dirac equation.

The problem of the singular modes is related to the fact that the Dirac operator as well as any Hamilton operator for charged particles is not selfadjoint in the presence of the AB potential. This would cause many problems for the unitary evolution of quantum systems unless selfadjoint extensions of these Hamilton operators exist. The selfadjoint extension procedure applied to this problem gives results which can be obtained by the direct calculation of the normalization integrals (for a detailed discussion see \cite{pragm}). However, this procedure does not fix the extension parameters which determines the behavior of the wave function at the origin. This situation is not satisfactory from the physical point of view. It can be overcome in turning to better defined models \cite{Hagen}, in particular, to the model of a solenoid of a finite radius with the uniform magnetic field \cite{Audretsch96}.

From Eq.(\ref{mc2}) we obtain for $\rho_0\rightarrow 0$
\begin{equation}\label{mc0}
|l|-\beta+2\beta\,{A_1\over B_1}\,{\Phi(A_1+1,B_1+1;\beta)\over
\Phi(A_1,B_1;\beta)} \sim \nu{\alpha-\xi_l\over \alpha+\xi_l}
\end{equation}
where
\begin{equation}\label{alpha}
\alpha = \left({p_\perp \rho_0\over
2}\right)^{2\nu}{\Gamma(-\nu+1)\over \Gamma(\nu+1)}
\end{equation}
and
\begin{equation}\label{xi}
\xi_l : = {b_l\over a_l} \sim \alpha {{\nu-|l|+\beta-2\beta\,(A_1/
B_1)\;(\Phi(A_1+1,B_1+1;\beta)/\Phi(A_1,B_1;\beta))}\over
{\nu+|l|-\beta+2\beta\,(A_1/ B_1)\;(\Phi(A_1+1,B_1+1;\beta)
/\Phi(A_1,B_1;\beta)})}.
\end{equation}
The asymptotic behavior of $\xi_l$ for $\rho_0 \rightarrow 0$ depends on the behavior of the denominator in Eq.(\ref{xi}). Since $\nu>0$ we have $b_l\sim \rho_0^{2\nu}\rightarrow 0$ for $\rho_0\rightarrow 0$ unless the following equality is valid,
\begin{equation}\label{denom1}
\lim_{\rho_0\rightarrow\infty} D : =
\lim_{\rho_0\rightarrow\infty}\left(\nu+|l|-\beta + 2\beta\,{A_1\over B_1}\,{\Phi(A_1+1,B_1+1;\beta) \over\Phi(A_1,B_1;\beta)}\right) = 0\,.
\end{equation}
This can happen only if
\begin{equation} \label{denom2}
\nu+|l|-\beta=0 \quad {\rm and} \quad \lim_{\rho_0\rightarrow\infty}\,A_1 = |l|-\varepsilon l+1-\varepsilon=0\,.
\end{equation}
In this case the parameter $\xi_l$ goes to zero,
\begin{equation}
\xi_l \sim \rho_0^{2(\nu - 1)} \rightarrow 0 \quad {\rm at} \quad \rho_0
\rightarrow\infty\,,
\end{equation}
unless
\begin{equation}\label{nu<1}
\nu = |l-\phi| < 1.
\end{equation}
The Eqs.(\ref{denom2}) and (\ref{nu<1}) are fulfilled for the positron ($\varepsilon>0$) singular mode $l=N$ only. It means that $a_N=0,\;b_N\neq 0,$ and that accordingly the first positron component with $l=N$ is singular and the second component is regular at $\rho=0:$
\begin{equation}
R_1(\rho) = J_{-\delta}(p_\perp \rho)\,, \quad R_2(\rho) = J_{1-\delta}(p_\perp \rho)\,.
\end{equation}
In this case the interaction between positron magnetic moment and the string magnetic field is attractive for the first (upper) component of the wave function and it is repulsive for the second (down) component. For the electron ($\varepsilon<0$) solution all coefficients $b_l=0.$ It means that the singular electron mode $l=-N-1$ has the regular first component and the singular second component
\begin{equation}
R_1(\rho) = J_{1-\delta}(p_\perp \rho)\,,\quad R_2(\rho) = - J_{-\delta}(p_\perp \rho).
\end{equation}

Introducing new notations we can rewrite the positron and electron solutions of the Dirac equation in the presence of the magnetic string in the following final forms which are valid both for regular and singular modes,
\begin{eqnarray}\label{abrm}
R_1(\rho) &=& e^{i{\pi\over 2}|l-\varepsilon N|}\;J_{\nu_1}(p_{\perp}\rho),
\nonumber\\
R_2(\rho) &=& e^{i{\pi\over 2}|l-\varepsilon N|}\;\epsilon_{l-\varepsilon N} \;
J_{\nu_2}(p_{\perp}\rho) = -ie^{i{\pi\over 2}|l+1-\varepsilon N|}\;J_{\nu_2}(p_{\perp}\rho)
\end{eqnarray}
with
\begin{eqnarray}\label{order12}
\nu_1 :=&& \epsilon_{l-\varepsilon N}\;(l-\phi)\,, \quad
\nu_2 := \epsilon_{l-\varepsilon N}\;(l+1-\phi)\,, \\ \nonumber
&& \epsilon_{l-\varepsilon N} := \left\{\begin{array}{rl}
1 &\; {\rm if}\ l \geq\varepsilon N\\
-1 &\; {\rm if}\ l<\varepsilon N
\end{array}\right.\,.
\end{eqnarray}
These radial solutions satisfy the normalization conditions (\ref{Rnc}).

\bigskip

{\bf Redefinition of $l:$} The matrix elements of the QED processes (bremsstrahlung of electrons and positrons, pair production and annihilation) in the presence of the AB potential contain integrals over the products  $\overline\Psi(x)\gamma_{\mu}A_{\mu}(x)\Psi(x),$ $\overline\Psi_c(x)\gamma_{\mu}A_{\mu}(x)\Psi_c(x),$ $\overline\Psi(x)\gamma_{\mu}A_{\mu}(x)\Psi_c(x),$
$\overline\Psi_c(x)\gamma_{\mu}A_{\mu}(x)\Psi(x),$ correspondingly, where $\Psi(x)$ and $\Psi_c(x)$ are the electron and charge conjugate positron functions. The integer number $N$ disappears from all these matrix elements after the redefinition of the angular quantum number $l$, $l \rightarrow l+\varepsilon N.$ This means that the matrix elements of the QED processes are independent of the integer part of the magnetic flux in units of the magnetic quantum. One can foresee this fact beforehand since the Dirac equation (\ref{de}) with $\beta=N+\delta$ can be transformed to the Dirac equation with $\beta=\delta$ by means of the gauge transformation
\begin{eqnarray}\label{gt}
\Psi(x)&\rightarrow&\Psi'(x)=\exp(-i\varepsilon
N\varphi)\Psi(x)\,,\nonumber \\
eA_i(x) : ={\varepsilon\beta\over\rho}&\rightarrow& eA'_i(x) = eA_i(x)
- i\exp(i\varepsilon N\varphi)\nabla_i\exp(-i\varepsilon N\varphi) =
{\varepsilon(\beta - N)\over\rho}. \nonumber
\end{eqnarray}
All observable quantities are conserved under this gauge transformation.

\section{Scattering of scalar and Dirac particles by a solenoid of finite radius} \label{scat}

The expressions obtained above present the partial positron and electron wave functions in terms of cylindrical modes. These states do not describe outgoing particles with definite linear momenta at infinity. In order to calculate the cross section of QED processes we need the {\em positron and electron scattering wave functions}. In external fields there exist two independent exact solutions of the Dirac equation, $\Psi^{(\pm)}(\vec{p}\,; x),$ which behave at large distances like a plane wave (propagating in the direction $\vec{p}$ given by $p_x = p_{\perp} \cos\varphi_p, \; p_y = p_{\perp} \sin\varphi_p, \; p_z$) plus an outgoing or ingoing cylindrical wave, correspondingly. Because of the damping of the cylindrical waves at large distances we may use these superpositions instead of plane waves. To evaluate correctly the matrix elements in the presence of external fields we have to take wave functions for ingoing (outgoing) particles which contain outgoing (ingoing) cylindrical waves. We turn first to a discussion of scalar particles.

\subsection{Low energy scattering of scalar particles}\label{low}

The wave function for scalar particles can be found as the eigenfunction of the complete set of the commuting operators (\ref{co}):
\begin{equation} \label{spinlesswf}
\Psi (x) = {1\over 2\pi} e^{-iE_p t + ip_3z}\;\psi (\rho, \varphi),
\quad \psi(\rho, \varphi) = R_l (\rho)\;e^{il\varphi}.
\end{equation}
Its radial modes $R_l(\rho)$ obey the radial Klein--Gordon equation,
\begin{equation} \label{slradeq}
R_l''+{1\over\rho} R_l' - \left({l\over \rho}-e
A_{\varphi}\right)^2 R_l + p_\perp^2\;R_l = 0
\end{equation}
which is similar to the Eq.(\ref{reR1}) but does not contain the term $e(A_{\varphi}'+A_{\varphi}/\rho)$ which describe the spin-magnetic field interaction.

For realistic models the internal radial solutions are some regular functions $c_l\;R_l^{\rm int}(\rho).$ For model (\ref{solvp}) of a solenoid with constant magnetic field this is
\begin{eqnarray}\label{slintsol'}
R_l^{\rm int}(\rho) &=& c_l\;x^{|l|\over 2}\;e^{-{x\over
2}}\;\Phi(A,B;x), \quad x = {\beta\over\rho_0^2}\;\rho^2\,, \\ \nonumber
A_l &=& {|l|-l+1\over 2} - {p_\perp^2\rho_0^2\over 4\beta}, \quad B_l
= |l|+1. \nonumber
\end{eqnarray}
The external solutions are similar to the first line of Eqs.(\ref{Rs}). It is convenient to rewrite them as follows
\begin{eqnarray} \label{slRs}
R_l(\rho) &=& (a_l + b_l\;e^{-i \pi\nu})\;J_{\nu}(p_\perp \rho) +
i\;b_l \sin\pi\nu\;H^{(1)}_{\nu}(p_\perp \rho), \\ \nonumber
\nu &=& |l-\phi|
\end{eqnarray}
where $H^{(1)}_{\nu}(x)$ are the Hankel function.
\noindent
The corresponding scattering wave functions can be obtained by the superposition of these cylindrical modes.

Since the plane wave term of the scattering wave function at $\rho\rightarrow\infty$ is defined only by the external solution, the scattering wave function can be written as follows
\begin{eqnarray}\label{frwf}
&&\Psi (\vec{p}\,; x) = {1\over 2\pi}\;e^{-iE_p
t+ip_3z}\;\psi (\vec{p}\,; \rho, \varphi), \\
&&\psi (\vec{p}\,; \rho, \varphi)=\sum_l {\tilde c}_{l-\varepsilon N}(\varphi_p) \;e^{i{\pi\over 2}|l-\varepsilon N|}\;\left[J_{\nu}(p_\perp \rho) - \mu_{l-\varepsilon N}(\varphi_p)\;H^{(1)}_{\nu}(p_\perp \rho) \right] \;e^{il\varphi}= \nonumber \\
&&e^{i\varepsilon N (\varphi-\varphi_p-\pi)}\sum_l e^{i{\pi\over 2}|l|}\;e^{i{\pi\over 2}(|l|-|l-\varepsilon\delta|)} \left[J_{|l-\varepsilon\delta|}(p_\perp \rho) - \mu_l(\varphi_p) \;H^{(1)}_{|l-\varepsilon\delta|}(p_\perp \rho)\right]\; e^{il(\varphi-\varphi_p)} \nonumber
\end{eqnarray}
with the coefficients
\begin{equation}\label{tildec}
{\tilde c}_{l-\varepsilon N}(\varphi_p) : = e^{-i\varepsilon N
\pi-il\varphi_p} \;e^{\pm i{\pi\over 2}(|l-\varepsilon N|-\nu)}
\end{equation}
The wave function (\ref{frwf}) behaves asymptotically as follows
\begin{equation}\label{asym}
\psi (\vec{p}\,;\rho,\varphi) \sim e^{i\varepsilon N (\varphi-\varphi_p-\pi)} \times\left[e^{ip_\perp\rho\cos(\varphi-\varphi_p)}
+f(\varphi -\varphi_p)\;{e^{\pm i p_\perp\rho}\over
\sqrt{\rho}}\right]
\end{equation}
where
\begin{equation}\label{f}
f (\varphi-\varphi_p) = {e^{-i{\pi\over 4}}\over\sqrt{2\pi p_\perp}}
e^{-i\varepsilon N (\varphi-\varphi_p)} \sum_l
\left[e^{i\pi(|l-\varepsilon N|-\nu)}\left[1-2\mu_{l-\varepsilon N}(\varphi_p) \right]-1\right] e^{il(\varphi-\varphi_p)}
\end{equation}
is the scattering amplitude. We note that the distorting factor $e^{i\varepsilon N (\varphi-\varphi_p-\pi)}$ appears in Eqs.(\ref{asym}) and (\ref{f}) because the vector potential (\ref{solvp}) decreases slowly at infinity.

The wave function (\ref{frwf}) and the scattering amplitude (\ref{f}) contain an arbitrary coefficient $\mu_l(\varphi_p)$ connected with the coefficients $a_l$ and $b_l$ of Eq.(\ref{slRs}). It describes how the finite radius $\rho_0$ of the solenoid influences the scattering cross section and can be found from the corresponding matching condition:
\begin{equation}\label{mu}
\mu_{l-\varepsilon N}(\varphi_p)
={J_{\nu}(p_\perp \rho_0) \over H_{\nu}^{(1)}(p_\perp \rho_0)}\;{(\ln J_{\nu}(p_\perp \rho_0))' - (\ln R_l^{\rm int}(p_\perp \rho_0))' \over (\ln H_{\nu}^{(1)}(p_\perp \rho_0))' - (\ln R_l^{\rm int}(p_\perp \rho_0))'}\,.
\end{equation}

The partial wave decomposition is the effective method for the investigation of low energy scattering. In this case the de Broglie wavelength of scattered particles is large compared with the size of the target ($\rho_0$), and the lower angular momenta contributes mainly to the scattering cross section. One can see from Eq.(\ref{mu}) that $\mu_l \rightarrow 0$ at $\rho_0\rightarrow 0.$  For any realistic model, including the model (\ref{solvp}), the internal solution is regular at $\rho_0 = 0.$ Then we have
\begin{equation}\label{L}
{{R'}_l^{\rm int}(p_\perp \rho_0)) \over R_l^{\rm int}(p_\perp \rho_0))} \sim {1\over p_\perp \rho_0}, \quad
L_{l-\varepsilon N} = \lim_{\rho_0\rightarrow 0}
{(\ln J_{\nu}(p_\perp \rho_0))' - (\ln R_l^{\rm int}(p_\perp \rho_0))' \over (\ln H_{\nu}^{(1)}(p_\perp \rho_0))' - (\ln R_l^{\rm int}(p_\perp \rho_0))'}
\sim {\rm const}
\end{equation}
and the second term in Eq.(\ref{frwf}) disappears completely at $\rho_0 \rightarrow 0,$
\begin{equation}\label{muL}
\mu_{l-\varepsilon N}(\varphi_p) \sim
i\sin\pi\nu\,{\Gamma(-\nu+1)\over
\Gamma(\nu+1)}\;L_{l-\varepsilon N}\,\left({p_\perp\rho_0\over
2}\right)^{2\nu} \rightarrow 0 ,  \end{equation}
with
\begin{equation} \label{L'}
L_{l-\varepsilon N} = {\nu-|l|+\beta-2\beta\eta_l \over
-\nu-|l|+\beta-2\beta\eta_l},\quad \eta_l={\Phi'(A_1,
B_1;\;\beta) \over \Phi(A_1, B_1;\;\beta)}\,.
\end{equation}
Accordingly we have obtained the following result: At low energies, $p_\perp\rho_0\rightarrow 0,$ the scattering wave function (\ref{frwf}) coincides with the AB wave function which behaves asymptotically as follows
\begin{equation}\label{corasym}
\psi (\vec{p}\,;\rho,\varphi)\sim e^{i\varepsilon N(\varphi-\varphi_p-\pi)} \left[e^{i\varepsilon \delta (\varphi-\varphi_p-\pi)}\; e^{ip_\perp\rho\cos(\varphi-\varphi_p)} + f_{AB} (\varphi -\varphi_p)\; {e^{ip_\perp\rho}\over \sqrt{\rho}}\right]
\end{equation}
in whole space outside a narrow region around the forward direction. The corresponding AB scattering amplitude reads
\begin{equation}\label{fab}
f_{AB} (\varphi-\varphi_p) = -\varepsilon \;{1\over\sqrt{2\pi p_\perp}}\;e^{-i{\pi\over 4}} \;e^{i\varepsilon {\varphi-\varphi_p\over 2}}\;{\sin\pi\delta\over\sin{\varphi-\varphi_p\over 2}}
\end{equation}
and the differential cross section of the AB scattering \cite{Aharonov59} is equal to
\begin{equation} \label{dcs}
{d\sigma\over d\varphi} = {p_\perp \over p}|f_{AB} (\varphi -\varphi_p)|^2
= {1\over 2\pi p}\;{\sin^2\pi\delta \over \sin^2{\varphi-\varphi_p\over 2}}.
\end{equation}

To find the first correction to the AB scattering amplitude at small $\rho_0$ one need to take a minimal value of $\nu.$ It is $\delta$ at $0<\delta<1/2,\;l=\varepsilon N$ and $1-\delta$ at $1/2<\delta<1,\; l=\varepsilon (N+1).$ We obtain
\begin{eqnarray}
\mu_0(\varphi_p) &=& i\sin\pi\delta\;{\Gamma(1-\delta)\over \Gamma(1+\delta)} \;L_0 \left({p_\perp\rho_0\over 2}\right)^{2\delta}, \nonumber \\
\mu_{\varepsilon}(\varphi_p) &=& i\sin\pi\delta\;{\Gamma(\delta)\over
\Gamma(2-\delta)} \;L_{\varepsilon} \left({p_\perp\rho_0\over 2}\right)^{2(1-\delta)}
\end{eqnarray}
with
\begin{equation}
L_0=1-{\delta\over (N+\delta)\eta_{\varepsilon N}}, \quad  L_{\varepsilon}={1\over
1+{1-\delta\over N+\delta}\eta^{-1}_{\varepsilon (N+1)}}.
\end{equation}

Then the scattering wave function reads as follows
\begin{eqnarray}\label{1approx}
\psi (\vec{p}\,;\rho,\varphi) &=& \psi_{AB} (\vec{p}\,; \rho, \varphi) +
\delta\psi (\vec{p}\,; \rho, \varphi),  \nonumber \\
\delta\psi (\vec{p}\,; \rho, \varphi) &=&
-\mu_0(\varphi_p)\;e^{-i{\pi\over 2}\delta}\;H^{(1)}_{\delta}(p_\perp\rho_0)\;
e^{i\varepsilon N(\varphi-\varphi_p-\pi)} \nonumber \\ &-& \mu_{\varepsilon}(\varphi_p)\;e^{-i{\pi\over 2}(1-\delta)} \;H^{(1)}_{1-\delta}(p_\perp\rho_0)\; e^{i\varepsilon (N+1)(\varphi-\varphi_p-\pi)}
\end{eqnarray}
where $\psi_{AB} (\vec{p}\,; \rho, \varphi)$ is the AB scattering wave function with the asymptotic behavior (\ref{corasym}). Now the additional term changes the modes with $l=\varepsilon N,\,\varepsilon (N+1)$ of the scattering amplitude $f_{AB}(\varphi-\varphi_p)$ given by Eq.(\ref{fab})
\begin{eqnarray}\label{frf}
f(\varphi-\varphi_p) &=& f_{AB}(\varphi-\varphi_p) +
\delta f(\varphi-\varphi_p) \nonumber \\
\delta f(\varphi-\varphi_p) &=& -{2\over \sqrt{2\pi
p_\perp}}\;e^{-i{\pi\over 4}}\left(\mu_0\;e^{-i\pi\delta}
+\mu_{\varepsilon}\;e^{-i\pi(1-\delta)} \;e^{i\varepsilon (\varphi-\varphi_p - \pi)}\right).
\end{eqnarray}
In this approximation the scattering cross section is equal to:
\begin{eqnarray} \label{frdcs}
{d\sigma\over d\varphi} &=& {p_\perp \over p}|f (\varphi
-\varphi_p)|^2 = {1\over 2\pi p}\;{\sin^2\pi\delta \over
\sin^2{\varphi- \varphi_p\over 2}} (1+\Delta), \nonumber \\ \Delta
&=& 4\sin{\varphi-\varphi_p \over 2} \sin({\varphi-\varphi_p \over 2}
+\pi\varepsilon\delta) \nonumber \\
&&\left[{\Gamma(1-\delta)\over \Gamma(1+\delta)}\; L_0
\left({p_\perp\rho_0\over 2}\right)^{2\delta} + {\Gamma(\delta)\over
\Gamma(2-\delta)} \;L_1 \left({p_\perp\rho_0\over
2}\right)^{2(1-\delta)} \right]
\end{eqnarray}
whereby $\Delta=0$ for $\rho_0=0.$

\subsection{High energy scattering of scalar particles}\label{high}

The partial wave method becomes ineffective when applied to the scattering of high energy particles. The smaller the de Broglie wavelength of scattered particles, the  higher angular momenta contribute to the scattering cross section. At $p_\perp\rho_0\rightarrow \infty$ this corresponds to $l\rightarrow \infty.$ In this case the scattering process can be expected to be the quasiclassical one and the WKB method \cite{Newton} can be used to obtain the scattering cross section.

The quasiclassical radial wave function for the model (\ref{solvp}) is equal to
\begin{equation}\label{wkb}
\psi^{WKB} = {\rm const}\;F^{-{1\over 4}}_l(\rho)\,\exp\left[i \int d\rho' \sqrt{F_l(\rho')}\right], \quad F_l(\rho) = p_\perp^2 - {(l-e\rho A_{\varphi})^2}\,.
\end{equation}
Then the phase shift in the WKB approximation is given by
\begin{equation}\label{shift}
\delta_l^{WKB} = {1\over 2}\pi l - \int_{\rho_{\rm min}}^{\infty} d\rho' \left[\sqrt{F_l(\rho)} - p_\perp\right] - p_\perp \rho_{\rm min}
\end{equation}
where we introduced the radius of the orbit in the uniform magnetic field $R=p_\perp / eB.$ $\rho_{\rm min}$ is the classical turning point of the motion inside the solenoid where the radial kinetic energy $F_l(\rho)$ goes to zero. $F_l(\rho)$ contains $\rho_0$ via $A_{\varphi},$ and in the zero radius limit Eq.(\ref{shift}) agrees with the corresponding expression for AB phase shift.

The classical deflection angle for charged particle moving in the magnetic field of the solenoid of finite radius (\ref{solvp}) can be obtained from the corresponding radial Hamilton--Jacobi equation. The WKB method gives the same result. Differentiating the phase shift (\ref{shift}) with respect to $l$ we obtain the classical deflection angle
\begin{equation}\label{wkbangle}
\varphi = 2{d\over dl}\delta^{WKB}_l = \pi - 2\int_{\rho_{\rm min}}^{\infty} {d\rho\over \rho} {\left(l- {p_\perp\rho^2\over 2R}\right) \over \sqrt{p_\perp^2\rho^2- \left(l - {p_\perp\rho^2 \over 2R}\right)^2}} - 2\int_{\rho_{\rm min}}^{\infty} {d\rho\over \rho} {\left(l- {p_\perp\rho_0^2\over 2R}\right) \over \sqrt{p_\perp^2\rho^2- \left(l - {p_\perp\rho_0^2 \over 2R}\right)^2}}\,.
\end{equation}
Introducing the impact parameter
\begin{equation}\label{impact}
a := {l\over p_\perp} - {\rho_0^2 \over 2R}
\end{equation}
and calculating the integral (\ref{wkbangle}) we obtain
\begin{equation}\label{wkbtraj}
\tan {\varphi\over 2} = {\sqrt{\rho_0^2 - a^2}\over R + a}.
\end{equation}
One can see from Eq.(\ref{wkbtraj}) that one or two values of the deflection angle correspond to one and the same value of the impact parameter $a.$ In this case classically unobtainable interference effects arise in the framework of the WKB approximation. Since the difference between the two impact parameters that lead to the same deflection angle is of macroscopic size in comparing with the particle wave length these interference effects are unobservable and can be neglected.

Then we obtain that the differential scattering cross section of scalar particles by the magnetic tube of the finite radius $\rho_0$ in the WKB approximation is equal to the classical cross section
\begin{equation}\label{wkbcs}
{d\sigma (\varphi)\over d\varphi} = \left|{da\over d\varphi}\right| =
\left\{ \begin{array}{ll}
{1\over 2}\sin{\varphi\over 2}{\left(\sqrt{\rho_0^2-R^2\sin^2{\varphi\over 2}} +R\cos^2{\varphi\over 2}\right)^2 \over \sqrt{\rho_0^2-R^2\sin^2{\varphi\over 2}}} & \mbox{at $R<\rho_0$} \\ \\
\sin{\varphi\over 2}{\rho_0^2-R^2\sin^2{\varphi\over 2} +R^2\cos^2{\varphi\over 2} \over \sqrt{\rho_0^2-R^2\sin^2{\varphi\over 2}}}    &  \mbox{at $R>\rho_0$}
\end{array}\,. \right.
\end{equation}
Integrating over $\varphi$ we find that the total scattering cross section of high energy incident particles is equal to $\sigma = 2\rho_0.$ This means that the scattering cross section decreases with the energy of incident particles approaching its geometrical value.

\subsection{Low energy scattering of Dirac particles}\label{dfinite}

In this section we consider the scattering of Dirac particles of low energies by the solenoid of the finite radius $\rho_0.$ As the initial approximation at $\rho_0 \rightarrow 0$ we prefer to use the radial function in the form of Eqs. (\ref{abrm}), (\ref{order12}) instead of Eqs.(\ref{Rs}). With this choice we avoid the problem connected with the specific behavior of the parameter $\xi_l$ of Eq.(\ref{xi}) at $\rho_0 \rightarrow 0$ with $l=N$ for the positron states and $l=-N-1$ for the electron states.

Then, for the solenoid of the finite radius, the external positron and electron radial solutions read as follows
\begin{eqnarray}\label{rsfr}
R_1(\rho) &=& e^{i{\pi\over 2}|l-\varepsilon N|}\left[J_{\nu_1}(p_{\perp}\rho) -\mu_{l-\varepsilon N}\;H^{(1)}_{\nu_1}(p_{\perp}\rho)\right], \\ \nonumber
R_2(\rho) &=& -ie^{i{\pi\over 2}|l+1-\varepsilon N|}
\left[J_{\nu_2}(p_{\perp}\rho) - \mu_{l-\varepsilon N}\;
H^{(1)}_{\nu_2}(p_{\perp}\rho)\right]
\end{eqnarray}
with the orders of the Bessel functions given by Eq.(\ref{order12}). The coefficients $\mu_{l}$ are defined by the corresponding matching conditions and decrease rapidly for $\rho_0\rightarrow 0.$ With these radial functions the cylindrical modes for the Dirac equation (\ref{de}) are given by Eqs.(\ref{ds}) and (\ref{uv}) with $R_3=R_1, R_4=R_2$ and the spin coefficients (\ref{c1c2'}), (\ref{c3c4'}) or (\ref{c1c2''}), (\ref{c3c4''}) correspondingly.

The scattering wave functions for the Dirac equation (\ref{de}),
\begin{equation} \label{dscatwf}
\psi_D(\vec{p}\,; \rho,\varphi) = \pmatrix{ u(\vec{p}\,; \rho,\varphi) \cr
v(\vec{p}\,; \rho,\varphi)\cr}
\end{equation}
are defined by the series
\begin{equation}\label{dscatu}
u(\vec{p}\,;\rho,\varphi) = \sum_l c_{l-\varepsilon N}\; \pmatrix{C_1\;R_1(\rho)\;e^{il\varphi} \cr
C_2\;R_2(\rho)\;e^{i(l+1)\varphi}\cr }
= \pmatrix{C_1\;\psi_1(\vec{p}\,; \rho,\varphi) \cr
-iC_2\;e^{i\varphi_p }\;\psi_2(\vec{p}\,; \rho,\varphi)\cr},
\end{equation}
\begin{equation}\label{dscatv}
v(\vec{p}\,;\rho,\varphi) = \sum_l c_{l-\varepsilon N}\; \pmatrix{C_3\;R_1(\rho)\;e^{il\varphi }\cr
C_4\;R_2(\rho)\;e^{i(l+1)\varphi }\cr}
= \pmatrix{C_3\;\psi_1(\vec{p}\,; \rho,\varphi) \cr
-iC_4\;e^{i\varphi_p }\;\psi_2(\vec{p}\,; \rho,\varphi)\cr}
\end{equation}
with the coefficients
\begin{equation} \label{dscatc}
c_{l-\varepsilon N}(\varphi_p)=e^{-i\varepsilon N \pi-il\varphi_p}
\;e^{i{\pi\over 2}\varepsilon\,\epsilon_{l-\varepsilon N}\,\delta}
\end{equation}
These coefficients differ from the coefficients ${\tilde c}_{l-\varepsilon N}(\varphi_p)$ (\ref{tildec}) of the scattering wave function for the scalar particles (\ref{frwf}) only at $l=N$ for the positron mode and at $l=-N-1$ for the electron mode.

Using Eqs.(\ref{rsfr}) we find the components of the Dirac scattering wave function for the positron solution
\begin{eqnarray} \label{dscatp}
\psi_1(\vec{p}\,;\rho,\varphi) &=& \psi(\vec{p}\,;\rho,\varphi)
+e^{iN(\varphi-\varphi_p-\pi)}\;i\sin\pi\delta\;e^{i{\pi\over 2}\delta}\; H^{(1)}_{\delta}(p_\perp \rho) + \delta\psi_1(\vec{p}\,;\rho,\varphi), \nonumber \\
\psi_2(\vec{p}\,; \rho, \varphi) &=& \psi(\vec{p}\,; \rho, \varphi)
+ \delta\psi_2(\vec{p}\,; \rho, \varphi)
\end{eqnarray}
and for the electron solution
\begin{eqnarray} \label{dscate}
\psi_1(\vec{p}\,;\rho,\varphi) &=& \psi(\vec{p}\,;\rho,\varphi) +
\delta\psi_1(\vec{p}\,;\rho,\varphi), \\
\psi_2 (\vec{p}\,;\rho,\varphi) &=&\psi(\vec{p}\,;\rho,\varphi)
+e^{-iN(\varphi-\varphi_p-\pi)}\;i\sin\pi\delta\;e^{i{\pi\over 2}\delta}\; H^{(1)}_{\delta}(p_\perp \rho)+\delta\psi_2(\vec{p}\,; \rho, \varphi)
\nonumber \end{eqnarray}
Here the function $\psi(\vec{p}\,; \rho, \varphi)$ is the scattering wave function (\ref{frwf}) for scalar particles, and the terms of the first approximation are equal to
\begin{eqnarray} \label{deltap}
\delta\psi_1(\vec{p}\,; \rho, \varphi) &=&
 - e^{iN(\varphi-\varphi_p-\pi)} \sum_l e^{i{\pi\over 2}|l|}\;
e^{i{\pi\over 2} \epsilon_l\,\delta}\,
\mu_l\,H^{(1)}_{\epsilon_l(l-\delta)}(p_\perp\rho)
\;e^{il(\varphi-\varphi_p)}, \\ \nonumber
\delta\psi_2(\vec{p}\,; \rho, \varphi) &=& - e^{iN(\varphi-\varphi_p-\pi)}
\sum_l e^{i{\pi\over 2}|l+1|}\;e^{i{\pi\over 2}(|l+1|-|l+1-\delta|)}\,
\mu_l\,H^{(1)}_{|l+1-\delta|}(p_\perp\rho)\;e^{i(l+1)(\varphi-\varphi_p)}
\end{eqnarray}
for the positron solution, and
\begin{eqnarray} \label{deltae}
\delta\psi_1(\vec{p}\,; \rho, \varphi) &=& - e^{-iN(\varphi-\varphi_p-\pi)}
\sum_l e^{i{\pi\over 2}|l|}\;e^{i{\pi\over 2}(|l|-|l+\delta|)}\;
\mu_l\;H^{(1)}_{|l+\delta|}(p_\perp\rho)\;e^{il(\varphi-\varphi_p)}, \\ \nonumber
\delta\psi_2(\vec{p}\,; \rho, \varphi) &=& - e^{-iN(\varphi-\varphi_p-\pi)}
\sum_l e^{i{\pi\over 2}|l+1|}\;e^{-i{\pi\over 2}\epsilon_l\delta}\;
\mu_l\;H^{(1)}_{\epsilon_l(l+1+\delta)}(p_\perp\rho)\;
e^{i(l+1)(\varphi-\varphi_p)}
\end{eqnarray}
for the electron solution. These terms decrease rapidly for $\rho_0\rightarrow 0.$ However Eqs.(\ref{dscatp}) and (\ref{dscate}) contain the additional terms which appear in the AB scattering wave functions for the Dirac particles for $\rho_0\rightarrow 0$ and which are absent for scalar particles. They arise due to a distortion of the positron radial mode with $l=N$ and the electron radial mode with $l=-1-N$ caused by the interaction the magnetic momenta of the Dirac particles with the string magnetic fields. This interaction does not influence other modes which go to zero on the string.

At $p_\perp\rho\rightarrow\infty$, and taking into account Eqs.(\ref{corasym}), we obtain the asymptotic behavior of the Dirac scattering wave function (\ref{dscatwf}) - (\ref{dscate}),
\begin{eqnarray}\label{dcorasympsi}
\psi(\vec{p}\,;\rho,\varphi) &\sim&e^{i\varepsilon N (\varphi-\varphi_p-\pi)}
\left[e^{i\varepsilon \delta (\varphi-\varphi_p-\pi)}\;e^{i p_\perp \rho\cos(\varphi-\varphi_p)}\right. \nonumber \\
&&+ \left.F(\varphi-\varphi_p)\;G(\varphi-\varphi_p)\; {e^{ip_\perp\rho}\over\sqrt{\rho}} \right]
\left(\begin{array}{c}
\displaystyle
u(\vec{p})\\
\displaystyle
v(\vec{p})
\end{array}\right)
\end{eqnarray}
where the scattering amplitude matrix $F(\varphi-\varphi_p)\; G(\varphi-\varphi_p)$ is defined by the expressions
\begin{eqnarray}\label{dampl}
F(\varphi-\varphi_p)&=&f_{AB}(\varphi-\varphi_p)\; e^{-i\varepsilon{\varphi-\varphi_p\over 2}}
+ \delta f(\varphi-\varphi_p)\;e^{i{\varphi-\varphi_p\over 2}}\,,\nonumber \\
\delta f(\varphi-\varphi_p)&=&\sqrt{2\over \pi p_\perp}\;e^{-i{\pi\over 4}}\;
\sum_l e^{i\pi\varepsilon\epsilon_l\delta}\;\mu_l\; e^{il(\varphi-\varphi_p)}
\end{eqnarray}
and
\begin{equation}\label{G}
G(\varphi-\varphi_p) =
\pmatrix{D(\varphi-\varphi_p)  &  0 \cr
      0 & D(\varphi-\varphi_p) \cr}
\end{equation}
with
\begin{equation}\label{F}
D(\varphi-\varphi_p) =
\pmatrix{ e^{-i{\varphi-\varphi_p\over 2}} &  0 \cr
      0 & e^{i{\varphi-\varphi_p\over 2}} \cr} \nonumber \\
= \cos{\varphi-\varphi_p\over 2} -i\sigma_3\; \sin{\varphi-\varphi_p\over 2}.
\end{equation}
The coefficients $\mu_l$ in Eq.(\ref{dampl}) are defined by the matching conditions and go to zero at $\rho_0\rightarrow 0.$

Here the plane wave bispinor
\begin{equation}\label{bispin}
\psi(\vec{p}) = \left(\begin{array}{c}
\displaystyle
u(\vec{p})\\
\displaystyle
v(\vec{p})
\end{array}\right) \,, \quad
u(\vec{p}) =
\left(\begin{array}{c}
\displaystyle
C_1\\
\displaystyle
-iC_2\;e^{i\varphi_p}
\end{array}\right) \,, \quad
v(\vec{p}) =
\left(\begin{array}{c}
\displaystyle
C_3\\
\displaystyle
-iC_4\;e^{i\varphi_p}
\end{array}\right)
\end{equation}
describes the ingoing particle with the linear momentum $\vec{p}.$

The differential cross section of the scattering for the Dirac particles is equal to
\begin{equation} \label{ddcs}
{d\sigma\over d\varphi} = {\vec{\rho}\;\vec{\jmath}\over j_p}
=|F(\varphi-\varphi_p)|^2\;{\psi^{\dagger}(\vec{p})\;
G^{\dagger}(\varphi-\varphi_p)\;\alpha_{\rho}(\varphi)\;
G(\varphi-\varphi_p)\;\psi(\vec{p})\over \psi^{\dagger}(\vec{p})\;
\alpha_{\vec{p}}(\varphi)\;\psi(\vec{p})}
\end{equation}
where $\vec{\rho}\;\vec{\jmath}$ and $j_p$ are flux densities of the scattered cylindrical and the incident plane waves.

Since $G^{\dagger}(\varphi-\varphi_p)\;\alpha_{\rho}(\varphi)\; G(\varphi-\varphi_p) = \alpha_{\rho}(\varphi_p),$ and the relations
\begin{eqnarray}
\psi^{\dagger}(\vec{p})\;\alpha_{\rho}(\varphi_p)\;\psi(\vec{p}) &=&
{p_\perp \over E_p},  \nonumber \\
\psi^{\dagger}(\vec{p})\;\alpha_{\vec{p}}(\varphi_p)\;\psi(\vec{p})
&=&\psi^{\dagger}(\vec{p})\left({p_\perp \over p}\alpha_{\rho}(\varphi_p) +
{p_3 \over p}\alpha_z \right)\psi(\vec{p}) = {p \over E_p}
\end{eqnarray}
are valid both for the helicity and the transverse states, we find that the differential cross section of the scattering of positrons and electrons by the solenoid of the finite radius
\begin{equation}  \label{dcs2}
{d\sigma\over d\varphi} = {p_\perp \over
p}|F(\varphi-\varphi_p)|^2 \end{equation}
is independent of the spin polarization quantum number $s.$ It coincides with the differential cross section for scalar particles. This means that spin effects do not appear in the scattering of spin particles in given helicity or transverse polarization states which are conserved in this process.

For particles polarized along an arbitrary vector $\vec{n}$ and described by the density matrix
\begin{equation} \label{dmatrix}
{\cal R}(\vec{n}) = {1\over 2}(1 + \vec{\sigma}\cdot\vec{n})\,,
\end{equation}
the differential cross section can be calculated by the formula
\begin{equation}\label{polbeam1}
{d\sigma\over d\varphi} = {p_\perp \over p}\;|F(\varphi-\varphi_p)|^2 {\rm Tr} \;{\cal R}(\vec{n}')\;D(\varphi-\varphi_p)\;{\cal R}(\vec{n})\;D^{\dagger}(\varphi-\varphi_p)
\end{equation}
where $\vec{n}'$ is the direction of the polarization vector after the scattering.

Calculating the trace on the right hand side of Eq.(\ref{polbeam1}) we find the differential cross section for particle beams of the given polarization
\begin{equation} \label{polbeam2}
{d\sigma\over d\varphi} = {p_\perp \over
p}\;|F(\varphi-\varphi_p)|^2\; \cdot {1\over 2}\{(1+\vec{n}\vec{n}') +
(\vec{n}\vec{n}'- n_3 n_3')\;\cos(\varphi-\varphi_p) +
[\vec{n}\times\vec{n}']_3\; \sin(\varphi-\varphi_p)\}.
\end{equation}
For the case of the same polarizations of the scattered and the incident beams, i.e. at $\vec{n}=\vec{n}',$ we obtain
\begin{equation}\label{polbeam3}
{d\sigma\over d\varphi} = {p_\perp \over p}\;|F(\varphi-\varphi_p)|^2   \left(\cos^2{\varphi-\varphi_p\over 2}+ n^2_3\; \sin^2{\varphi-\varphi_p\over 2}\right).
\end{equation}
One can see that the spin effect occurs for polarized particles beams.

\section{Conclusion}\label{concl}

We studied in detail the Dirac equation in cylindrical magnetic fields of a fixed direction and found eigenfunctions of a complete set of commuting operators which consists of the Dirac operator itself, the $z$-components of linear and total angular momenta, and of one of two possible spin polarization operators. The spin structure of the solutions is independent on the radial distribution of the magnetic field and is fixed by one of the spin polarization operators. The magnetic field influences only the radial modes. We solved explicitly the radial equations for the model where the magnetic field is distributed uniformly inside a solenoid of a finite radius. From this realistic model we obtained correct solutions of the Dirac equation for the AB potential (magnetic string) in the zero radius limit. Also we considered carefully the scattering of scalar and Dirac particles by the magnetic tube of a finite radius. For low energy particles the scattering cross section coincides with the AB scattering cross section. In this case the de Broglie wavelength of particles is bigger compared with the tube radius and the influence of enclosed magnetic flux leads to the large cross section. This mechanism of interaction of quantum particles with the string magnetic field is more effective than the local interaction with the magnetic field.  We calculated the first order corrections connected with a finite radius of the tube. When the energy of incident particles increases the influence of the enclosed magnetic flux weakens and the cross section decreases. At high energies we obtained therefore the classical result for the scattering cross section.

The exact solutions derived above will serve as well as basis for future calculations. We plan to study how the finite size of the solenoid influences the differential cross sections of the bremsstrahlung emitted by an electron and the pair production by a single photon at different energies of the incoming particles.


\section*{Acknowledgments}

V.~S.~thanks Prof.J.~Audretsch and the members of his group for friendly atmosphere and hospitality at the University of Konstanz. This work was supported by the Deutsche Forschungsgemeinschaft and by Russian Foundation for Basic Research (96-02-16053-a).



\begin{thebibliography}{99}

\bibitem{Aharonov59}
Y.~Aharonov and D.~Bohm, Phys.~Rev.~{\bf 119}, 485 (1959).

\bibitem{Akhiezer65}
A.I.Akhiezer and V.B.Berestetskii,$\;$ {\it Quantum electrodynamics}, Interscience Publishers, New York (1965).

\bibitem{pragm}
J.Audretsch, U.Jasper and V.D.Skarzhinsky, J.~Phys.~A {28}, 2359, (1995).

\bibitem{Audretsch96}
J.Audretsch, U.Jasper, and V.D.Skarzhinsky, Phys.Rev.D, {\bf 53}, 2178, (1996).

\bibitem{Bagrov90}
V.G.Bagrov and D.M.Gitman, {\it Exact solutions of relativistic wave equations}, Kluwer (1990)

\bibitem{Landau79}
V.B.Berestetskii, E.M.Lifshitz and L.P.Pitaevskii$\;$ {\it Quantum
electrodynamics}, vol.4, Pergamon Press (1979).

\bibitem{Erber66}
T.Erber, Rev.Mod.Phys., {\bf 38}, 626, (1996).

\bibitem{Gradshteyn80}
I.S.Gradshteyn and I.M.Ryzhik, {\it Table of integrals, series and products}, Academic Press (1980).

\bibitem{Hagen}
C.R.Hagen,  Phys.Rev.Lett.{\bf 64}, 503 (1990); Int.J.Mod.Phys.A {\bf 6}, 3119 (1991); Phys.Rev.D, {\bf 52}, 2466, (1995).

\bibitem{Klepikov54}
N.P.Klepikov,  JETF (Sov.Phys.JETP), {\bf 26}, 19, (1954).

\bibitem{Newton}
R.G-Newton,$\;$ {\it Scattering theory of Waves and Particles},
Springer-Verlag,  Berlin (1982).

\bibitem{Peshkin89}
M.Peshkin and A.Tonomura,$\;$ {\it The Aharonov--Bohm effect},
Springer-Verlag,  Berlin (1989).

\bibitem{Ritus79}
V.I.Ritus, Journal of Soviet Laser Research {\bf 6}. No.5, Sept.Oct., 497 (1985) - translated from {\em Proc.P.N.Lebedev Phys.Inst.{\bf 111}} (in Russian),  ed.by N.G.Basov, Nauka, Moscow (1979); A.I.Nikishov,
Journal of Soviet Laser Research {\bf 6}. No.6, Nov.Dec., 619 (1985) -
translated from {\em Proc.P.N.Lebedev Phys.Inst.{\bf 111}} (in Russian),
ed.by N.G.Basov, Nauka, Moscow (1979).

\bibitem{Schwinger51}
J.Schwinger, Phys.Rev., {\bf 82}, 664, (1951).

\bibitem{Skarzhinsky96}
V.D.Skarzhinsky, J.Audretsch, and U.Jasper, Phys.Rev.D, {\bf 53}, 2190, (1996).

\bibitem{Sokolov68}
A.A.Sokolov and I.M.Ternov, {\it Synchrotron radiation}, Akademie-Verlag, Berlin (1968).

\bibitem{Volkov35}
D.M.Volkov,  Z.Phys., {\bf 94}, 250, (1935).

\end{thebibliography}
\end{document}